\documentclass[prc,twocolumn,pdflatex,nofootinbib,superscriptaddress]{revtex4-1}

\pdfoutput=1

\usepackage{graphicx}
\usepackage{bm,amsmath,amssymb,amsfonts}
\newcommand{\hsp}{\hspace*{1pt}}
\newcommand{\hspm}{\hspace*{.5pt}}
\newcommand{\ds}{\displaystyle}
\newcommand{\be}{\begin{equation}}
\newcommand{\ee}{\end{equation}}
\newcommand{\bel}[1]{\be\label{#1}}
\newcommand{\re}[1]{Eq.~(\ref{#1})}
\newcommand{\thickhata}[1]{\mathbf{\hat{\text{\hspace*{-2pt}$#1$}}}}
\newcommand{\thickhat}[1]{\mathbf{\hat{\text{$#1$}}}}
\usepackage{color}

\begin{document}

\title{Condensation of interacting scalar bosons at finite temperatures}

\author{I.~N.~Mishustin${}^{1,2}$,
D.~V.~Anchishkin${}^{1,3,4}$, L.~M.~Satarov${}^{1,2}$, O.~S.~Stashko${}^4$,
and~H. Stoecker${}^{1,5,6}$
\\
${}^1$ \textit{Frankfurt Institute for Advanced Studies, 60438 Frankfurt am Main, Germany}\hfill
\\
${}^2$ \textit{National Research Center "Kurchatov Institute", 123182 Moscow, Russia}\hfill
\\
${}^3$ \textit{Bogolyubov Institute for Theoretical Physics, 03143 Kyiv, Ukraine}
\\
${}^4$ \textit{Taras Shevchenko National University of Kyiv, 03022 Kyiv, Ukraine}
\\
${}^5$ \textit{Johann Wolfgang Goethe University, 60438 Frankfurt am Main, Germany}
\\
${}^6$ \textit{GSI Helmholtzzentrum f\"ur Schwerionenforschung GmbH, 64291 Frankfurt am Main, Germany}
}

\begin{abstract}
Thermodynamical properties of an interacting system {of scalar bosons} at finite
temperatures are studied within the framework of a field-theoretical model containing
the attractive and repulsive self-interaction terms. Self-consistency relations between
the {effective mass} and thermodynamic functions are derived in~the mean-field approximation. We show that for a sufficiently strong attractive interaction a first-order phase transition develops in the system via the formation of a~scalar condensate. An interesting prediction of this model is that the condensed phase appears within a~finite temperature interval and is characterized by a constant scalar density of  Bose particles.
\end{abstract}

\maketitle

\section{Introduction}

In recent years properties of hot and
dense hadronic matter have attracted considerable interest.
Such matter can be produced in relativistic nucleus-nucleus collisions which
are experimentally studied in many laboratories.
Both, QCD-motivated effective mo\-dels and lattice simulations indicate that the chiral
symmetry restoration and the deconfinement phase transition (PT) should occur
at high temperatures and particle densities. The
properties of hadrons will be strongly modified under such conditions.

The hadron resonance gas model with excluded volume corrections is widely used in the
literature (see e.g.~\mbox{\cite{andronic-2006,satarov-2009,vovchenko-anch-2015,vovchenko-2017,vovchenko-2017a,
anch-vovchenko-2015}})
to fit the lattice results and the experimental data from relativistic heavy-ion collisions. These calculations
show that the pion densities may reach the values \mbox{${n_{\pi}\sim}(0.1-0.2)~\textrm{fm}^{-3}$} at
tempera\-tures~\mbox{${T\sim}(140-160)~\textrm{MeV}$}. At such high densities the interaction effects became important. Dense hadronic systems have been investigated recently
within the mean-field approach, using the van der Waals~\cite{vovchenko-2017a,anch-vovchenko-2015,anch-2016} and Skyrme-like \cite{anch-2019} models. The pion gas with repul\-sive~$\phi^4$~interaction has been considered earlier in Ref.~\cite{vos-2018}.

In the present paper we study properties of interac\-ting bosonic
systems.~This problem has been studied
previously, starting from the pioneer works of Migdal and coworkers
\cite{migdal-1972,migdal-1974,migdal-1978,saperstein-1990} and later
by many authors \mbox{using} different models and methods.
The formation of classical pion fields in relativistic nucleus-nucleus
collisions was discussed in
Refs.~\cite{anselm-1991,blaizot-1992,bjorken-1992}.
Pionic systems with a~\mbox{finite} isospin chemical potential were considered
within effective chiral models in Refs.~\cite{mishustin-greiner-1993,son-2001,kogut-2001,toublan-2001,mammarella-2015,carignano-2017}.
Such systems have been investigated also by
lattice QCD simula\-tions~\cite{brandt-2016,brandt-2017}.

Below we consider a general system of interacting
bosons associated with a scalar field $\phi$.
Following Ref.~\cite{anch-nazarenko-2006} we introduce an effective Lagrangian which contains
the attractive ($\phi^4$) and the repulsive ($\phi^6$) self-interaction terms.
The calculations are carried out within the mean-field approach,
taking into account the thermodynamic consistency conditions.
It is assumed that the system  has no conserved charge and, therefore,
it is characterized by a vanishing chemical potential.
It will be shown that at strong enough attrac\-tive interactions, the quasiparticle's effective
mass vanishes and the system undergoes a first-order PT with formation of a~scalar
Bose-Einstein condensate. This phase \mbox{exists} within a~\mbox{finite} temperature interval, determined by the
parameters of the interaction.

\section{Formulation of the model}

Let us consider a real (pseudo-)scalar field\,
$\thickhata{\phi}\hspm (x)$, which has no conserved charge.
The corresponding Lagrangian can be written as~(\mbox{$\hbar=c=1$})
\bel{lagr1}
 {\mathcal{L}}=\frac{1}{2} \left[\partial_\mu\,\thickhata{\phi}\hsp (x)\hsp
\partial^{\,\mu}\,\thickhata{\phi}\hsp (x)
- m^2\,\thickhat{\sigma}\hsp (x)\right]+\mathcal{L}_{\rm int}\hsp [\hsp\thickhat{\sigma}(x)]\,.
\ee
Here $\thickhat\sigma\hspm (x)=\thickhata{\phi}^{\hsp 2}\hspace*{-2pt}(x)$ is the 'scalar density' operator,  $m$~is the boson
mass in the vacuum, and~$\mathcal{L}_{\rm int}$ is the interaction Lagrangian. One can decompose $\mathcal{L}_{\rm int}$
in powers of \mbox{$\delta\hsp\thickhat{\sigma}=\thickhat{\sigma}-\sigma$} where $\sigma=\langle\thickhat{\sigma}\rangle$
is the mean value of the scalar density. Here and below, angular brackets denote
the statistical averaging in the grand-canonical ensemble:
\bel{densm}
\left\langle\thickhata{A}\right\rangle
=\frac{1}{Z} {\rm Tr}\left[ e^{- \beta\hspm \left(\hsp\thickhata{H} - \mu\,\thickhata{N} \right) }\, \thickhata{A} \right],
~~Z\, =\, {\rm Tr}\left[ e^{- \beta\hspm\left(\hsp\thickhata{H} - \mu\,\thickhata{N} \right) } \right] ,
\ee
where $\beta=1/T$ is the inverse temperature, $\mu$ is the chemi\-cal potential, $\thickhata{H}$ and $\thickhata{N}$
are the Hamiltonian and the particle number operators, respectively. Below we consider only the case $\mu=0$
which corresponds to a system, where the total particle number is
not conserved, but is determined by the temperature (like photons or \mbox{$\pi$-mesons}). However, this {model} can be easily extended to nonzero~$\mu$\,.

In the following we apply the mean-field approxima\-tion~(MFA), where
fluctuations of the scalar field up to the first order in \mbox{$\delta\thickhat{\sigma}$}  {(i.e., up to the second order
in~$\thickhat{\varphi}$)} are taken into account.
Then one can write
\bel{mfa}
\mathcal{L}_{\rm int}\hspm (\thickhat{\sigma})\simeq \mathcal{L}_{\rm int}\hspm (\sigma)+
\delta\hspm\thickhat{\sigma}\hsp\mathcal{L}_{\rm int}^\prime\hspm (\sigma)~~~~(\textrm{MFA})\,,
\ee
where the prime denotes the differentiation with respect to~$\sigma$\hspm.
The effective Lagrangian  {of field excitations} can be represented as
\bel{mfal}
 {\mathcal{L}}\simeq
\frac{1}{2}\left[\partial_\mu\,\,\thickhata{\phi}(x)\hsp\partial^{\,\mu}\,\thickhata{\phi}(x)
- M^2(\sigma)\,\,\thickhata{\phi}^{\,2}(x)\right]+P_{\rm ex}(\sigma)\,,
\ee
where Eqs.~(\ref{lagr1}) and~(\ref{mfa}) have been used. The quantity
\bel{efma}
M^2(\sigma)=m^2+\Pi (\sigma),
\ee
is the effective mass of bosonic quasiparticles and
\mbox{$\Pi (\sigma)=-2\hspm\mathcal{L}_{\rm int}^\prime\hspm (\sigma)$}
is the 'polarization operator'. The last term in~\re{mfal},
\bel{pex1}
P_{\rm ex}(\sigma)=\mathcal{L}_{\rm int}\hspm (\sigma)-
\sigma\mathcal{L}_{\rm int}^\prime\hspm (\sigma)
\ee
is the so-called 'excess' pressure. It will be shown that this term
gives rise to the pressure shift due to interactions.
Note that the argument $\sigma$ in the functions $M,\Pi,P_{\rm ex}$ is a $c$--number.
In fact, \re{efma} can be regarded as the 'gap' equation for $M=M\hspm (\sigma)$, as
$\sigma$ itself is a function of $M$ (see below).

It follows from \re{mfal} that the equation of motion for the operator~~$\thickhata{\phi}$
can be written in Klein-Gordon form with the effective mass $M\hspm (\sigma)$
\bel{eofm}
\partial^{\,\mu}\partial_\mu\,\,\thickhata{\phi}\hspm + M^2(\sigma)\,\,\thickhata{\phi}=0\,.
\ee
Equations (\ref{efma}) and (\ref{pex1}) lead to the following self-consistency relation
\bel{pex2}
\frac{\ds d\hspm P_{\rm ex}}{\ds d\hspm\sigma}=\frac{\sigma}{2}\frac{\ds d\hspm\Pi}
{\ds d\hspm\sigma}\,,
\ee
which should hold for any $\mathcal{L}_{\rm int}$. This is another form of the
corresponding relation derived in~Refs.~\cite{anch-1992,anchsu-1995}.
The shift of a single-particle energy (commonly called as the effective potential)
in nonrelativistic limit is equal to $U\simeq\Pi/(2m)$ and $\sigma\simeq n/m$, where $n$ is the particle number density.

\section{Derivation of the thermodynamic functions}

Let us now introduce the momentum operator \mbox{$\thickhat{\pi}(x)=\partial_t\,\,\thickhata{\phi}(x)$} which
satisfies the equal-time commutation relation
\bel{comr}
\left[\,\thickhata{\phi}\hspm (t,\bm r),\hsp\thickhat{\pi}\hspm (t,\bm r') \right]=i \delta^3(\bm r-\bm r')\,.
\ee
In the MFA, the Hamiltonian density operator \mbox{$\thickhata{\mathcal{H}}=\thickhat{\pi}\hsp\partial_t\,\,\thickhata{\phi}-\mathcal{L}$} takes the form
\bel{eq:hamiltonian1-real}
\thickhata{\mathcal{H}}\simeq
\frac{1}{2}\left[\thickhat{\pi}^2(x)+\bm\nabla\,\thickhata{\phi}\hspm (x)\cdot\bm\nabla\,\thickhata{\phi}\hspm (x)\,+ M^2(\sigma)\,\,\thickhata{\phi}^{\,2}(x)\right]-P_{\rm ex}(\sigma)\,.
\ee
Using solutions of the Klein-Gordon equation~(\ref{eofm}) one can represent the scalar field~~$\thickhata{\phi}\hsp (x)$ as
\bel{pwq}
\thickhata{\phi}\hsp (x)= {g}\int\frac{\ds d^{\,3}k}{\ds (2\pi)^3\sqrt{2\omega_{\bm k}}}\left[a_{\bm k}e^{-ik\cdot x}+
a^+_{\bm k}e^{ik\cdot x}\right].
\ee
Here $g$ is the spin-isospin statistical weight,
\bel{disp}
k^0=\omega_{\bm k}=\sqrt{{\bm k}^2+M^2(\sigma)},
\ee
and $a_{\bm k}, a^+_{\bm k}$ are the annihilation and creation operators, respectively.
They obey the  standard commutation relations:
\bel{crfa}
\left[a_{\bm k},a^+_{\bm{k}^\prime}\right]=(2\pi)^3\delta(\bm{k}-\bm{k}^\prime),
~~\left[a_{\bm k},a_{{\bm k}^\prime}\right]
=\left[a^+_{\bm k},a^+_{\bm{k}^\prime}\right]=0\,.
\ee

Substituting (\ref{pwq}) into (\ref{eq:hamiltonian1-real}) one gets the Hamiltonian operator
in the MFA
\bel{hamm}
\thickhata{H}=\mbox{$\ds\int\hspace*{-3pt}d^{\,3}x~\thickhata{\mathcal H}$}=V\left[\mbox{$ {g}\hspace*{-2pt}\ds\int\hspace*{-2pt}
\frac{\ds d^{\,3}k}{\ds (2\pi)^3}\,
\normalsize\omega_{\bm k}\hsp a^+_{\bm k}\hsp a_{\bm k}-P_{\rm ex}(\sigma)$}\right],
\ee
where $V$ is the total system's volume.
Using Eqs.~(\ref{densm}) and (\ref{hamm}) one can calculate all thermodynamic
functions of the considered system.
In the~MFA the equilibrium momentum distribution coincides with that
of an ideal gas of bosons with the effective mass~$M(\sigma)$,
\bel{bspe}
n_{\bm k}(\sigma)\equiv\langle a^+_{\bm k}\hsp a_{\bm k}
\rangle=(e^{\beta\hspm\omega_{\bm k}}-1)^{-1}\,,
\ee
where $\omega_{\bm k}$ is given by~\re{disp}.

The equation for the scalar density \mbox{$\sigma=\langle~\thickhata{\phi}^{\,2}\rangle$} is obtained by direct calculation using \mbox{Eqs.~(\ref{pwq}),~(\ref{bspe})}. This leads to the following gap equation
\bel{gape1}
\sigma=\sigma_{\rm th}(M,T)~~\textrm{where}~~\sigma_{\rm th}(M,T)=g\int
\frac{\ds d^{\,3}k}{\ds (2\pi)^3}\,\frac{n_{\bm k}(\sigma)}{\omega_{\bm k}}\,.
\ee
Note that the number density of quasiparticles, \mbox{$n=g\int \frac{d^{\,3}k}{(2\pi)^3}\,n_{\bm k}(\sigma)$}, does not contain $\omega_{\bm k}$ in the~denominator.
The pressure $P=\frac{T}{V}\ln{Z}$ is calculated as
\bel{prmf}
P=P_{\hspm\rm kin}(M,T)+P_{\rm ex}\hspm (\sigma)\,.
\ee
Here the first term is the pressure of an ideal gas of bosonic quasiparticles with mass $M$,
\bel{prmf1}
P_{\hspm\rm kin}(M,T)=
\frac{g}{3}\int \frac{\ds d^{\,3}k}{\ds (2\pi)^3}\frac{k^2}{\omega_{\bm k}}\,n_{\bm k}(\sigma),
\ee
where $n_{\bm k}(\sigma)$ is given in~\re{bspe}.
To obtain \mbox{$P=P\hspm (T)$} one should calculate $\sigma$ and $M$ as functions of $T$ by
simultaneously solving the system of equations (\ref{efma}) and~(\ref{gape1}).

Using Eqs.~(\ref{hamm}) and (\ref{bspe}) one can calculate the energy density $\varepsilon=\langle\,\thickhata{H}\rangle/V$ as
\bel{end1}
\varepsilon=g\int \frac{\ds d^{\,3}k}{\ds (2\pi)^3}\,
\omega_{\bm k}\,n_{\bm k}(\sigma)-P_{\rm ex}(\sigma)\,.
\ee
Within the MFA, the entropy density $s=(\varepsilon + P)/T$ formally coincides with that for the ideal gas of quasiparticles:
\bel{entd}
s=\frac{g}{T}\int \frac{\ds d^{\,3}k}{\ds (2\pi)^3}\hsp
\left(\omega_{\bm k}+\frac{k^2}{3\hsp\omega_{\bm k}}\right)n_{\bm k}(\sigma)\,.
\ee
Here the interaction effects enter via the effective mass~$M\hspm (\sigma)$\,.

Equations~(\ref{prmf}), (\ref{entd}) satisfy the condition of thermodynamic consistency, \mbox{$s=dP/dT$}\hspace*{1.5pt}\cite{gorenstein-1995}. Indeed, in accordance with~\re{prmf},
one has
\bel{dpdt}
\frac{dP}{dT}=\left(\frac{\partial P_{\rm kin}}
{\partial T}\right)_M+\left[\left(\frac{\partial P_{\rm kin}}{\partial M}\right)_T+
\frac{dP_{\rm ex}}{dM}\right]\hsp\frac{dM}{dT}=s\,.
\ee
Here it is taken into account that\hspm\footnote
{
The second equality in~\re{dpdt1} is obtained by direct diffe\-rentiation~of~(\ref{prmf1}), while
the last relation is obtained from~(\ref{pex2}).
}
\bel{dpdt1}
\left(\frac{\partial P_{\rm kin}}
{\partial T}\right)_M=s,~~\left(\frac{\partial P_{\rm kin}}{\partial M}\right)_T=-M\sigma_{\rm th},
~~\frac{dP_{\rm ex}}{dM}=M\sigma\,.
\ee
The terms in square brackets of \re{dpdt} add to \mbox{$M\hspm (\sigma-\sigma_{\rm th})$} which is zero
both in the normal phase~\mbox{$(\sigma=\sigma_{\rm th})$} as well as in the phase with a condensate, where \mbox{$M=0$}~(see below).

\section{Formation of scalar Bose condensate}

When the lowest energy level with $|\bm k| = 0$ is 'macroscopically' occupied, one should treat this level separately,
i.e. write instead of~(\ref{pwq}) the equation \mbox{$\thickhata{\phi}\hspm (x)=\varphi+\thickhat{\chi}(x)$}.
Here $\varphi$ is the classical part of the field operator~~$\thickhata \phi$, which
is ana\-logous to the Bose-Einstein condensate of massive particles. Below we denote such a~classical field as Scalar Bose Condensate (SBC). The
second term $\thickhat{\chi}$ is the fluctuating (quantum) part of~~$\thickhata \phi$, which sa\-tisfies the relation \mbox{$\langle\thickhat{\chi}\rangle=0$}. In~the domain with  {SBC}, $\varphi\neq 0$ and instead of~\re{gape1}, we now write
\bel{gape2}
\sigma\equiv\langle{~\thickhata{\phi}^{\,2}}\rangle=\sigma_{\rm cond}+\sigma_{\rm th}\,.
\ee
Here both, $\sigma_{\rm cond}=\varphi^2$ and $\sigma_{\rm th}=\langle{\thickhat{\chi}^2}\rangle$,
are nonzero, positive quantities.

By analogy to the case of a conserved charge (\mbox{$\mu\ne 0$}) we assume that the  {SBC} occurs for states where the occupation number $n_{\bm k}(\sigma)$ diverges at \mbox{$k\to 0$}.
As follows from~\re{bspe}, the~divergence takes place at
$M\hspm (\sigma)\to 0$\hspm\footnote
{
In Refs.~\mbox{\cite{migdal-1972,migdal-1974,migdal-1978,saperstein-1990}} the onset of
pion condensation was determined by the occurrence of solutions with $\omega^2_{\bm k}\leqslant 0$.
}.
This condition is satisfied at $\sigma=\sigma_0$ where $\sigma_0$ is the root of the equation:
\bel{cbec}
M^2(\sigma_0)=m^2+\Pi\hspm (\sigma_0)=0.
\ee
As we will see below, \re{cbec} gives a~ne\-cessary, but in general, not sufficient
condition for the  {SBC} formation. According to \re{cbec}, this condensation requires a~ne\-gative $\Pi\hspm (\sigma)$
which implies a sufficiently strong attractive interaction of particles.

For massless quasiparticles (\mbox{$\omega_{\bm k}=k$}) \re{gape1} yields
\bel{scdm}
\sigma_{\rm lim}(T)\equiv\hsp\sigma_{\rm th}(M=0,T)=\frac{gT^2}{12}\,.
\ee
Substituting (\ref{scdm}) into~(\ref{gape2}) gives
\bel{scbd}
\sigma_{\rm cond}(T)=\sigma_0-\frac{gT^2}{12}>0\,.
\ee
Hence, the  {SBC} may appear only at temperatures
\bel{cbec1}
T<T_{\rm max}=\sqrt{\frac{12\hspm\sigma_0}{g}},
\ee
where $\sigma_0$ is found by solving~\re{cbec}.

In the phase with  {SBC}, the thermodynamical
quantities $P\hspm (T),\varepsilon\hspm (T), s\hspm (T)$ and~$n\hspm (T)$
are given by the corresponding formulae in the preceding section
with \mbox{$M=0$}, \mbox{$\sigma=\sigma_0$}, namely
\begin{eqnarray}
&&P=\frac{g\hspm\pi^2}{90}T^4+P_{\rm ex}(\sigma_0),~~\varepsilon=\frac{g\hspm\pi^2}{30}T^4-
P_{\rm ex}(\sigma_0),~~\nonumber\\
&&s=\frac{2\hspm g\hspm\pi^2}{45}T^3=\frac{2\hspm\pi^4}{45\hsp\zeta(3)}\,n\simeq
3.60\hsp n\,,~~\label{bter}
\end{eqnarray}
where $\zeta(x)=\sum_{k=1}^\infty k^{-x}$ is the Riemann function.
Note that $-P_{\rm ex}(\sigma_0)$ plays the role of an effective 'bag
constant' for this phase.

\section{Bosonic system with Skyrme-like interaction}

The above results are valid in the MFA and
they do not depend on a specific form of the interaction.
For illustration, below we consider the Skyrme-like ($\phi^4-\phi^6$) interaction Lagrangian
\bel{skin}
\mathcal{L}_{\rm int}(\sigma)=\frac{a}{4}\hsp\sigma^2-\frac{b}{6}\hsp\sigma^3\,,
\ee
where $a,b$ are positive constants. The first and the \mbox{second} terms in (\ref{skin}) describe, respectively,
the attractive and the repulsive interactions between the scalar bosons. Note that the bosonic system
becomes unstable in the limit $b\to 0$ (see below). Using formulas of preceding section
one has
\begin{eqnarray}
&&\Pi\hspm (\sigma)=-2\mathcal{L}_{\rm int}^\prime=
-a\hspm\sigma +b\hspm\sigma^2\,,\label{skin1}\\
&&P_{\rm ex}\hspm (\sigma)=\mathcal{L}_{\rm int}-\sigma\mathcal{L}_{\rm int}^\prime=
-\frac{a}{4}\hspm\sigma^2 +\frac{b}{3}\hspm\sigma^3\,.\label{skin2}
\end{eqnarray}

The scalar density $\sigma$ of stable matter should satisfy the condition
\bel{stabc}
M^2(\sigma)=m^2-a\hspm\sigma+b\hsp\sigma^2\geqslant 0\,.
\ee
As discussed above, the onset of condensation corresponds to the equality sign in this expression.

It is convenient to introduce parameter~\mbox{$\kappa=a/(2m\sqrt{b})$} to consider differen possibilities predicted by the model. They are
presented  in Fig.~1. In the case of 'weak attraction', $\kappa<1$, the right hand side of~\re{stabc} is positive at any $\sigma$, i.e. no
{SBC} can be formed. In this case all thermodynamic quantities change smoothly with temperature and no PT is expected. Never\-theless, the
equation of state differs signi\-ficantly from the ideal gas with the vacuum masses of bosons. In the region $\kappa<1$,
the equilibrium values of~$\sigma$ and~$M$ as functions of $T$ are found
by simultaneously solving the equations
\bel{gape3}
M=\sqrt{m^2-a\hspm\sigma+b\hsp\sigma^2}\,,~~\sigma=\sigma_{\rm th}(M,T)\,,
\ee
where $\sigma_{\rm th}(M,T)$ is defined in~\re{gape1}. Then, the pressure $P(T)$ can be calculated by using Eqs.~(\ref{prmf}),~(\ref{prmf1}) and~(\ref{skin2}). It is an increasing function of $T$,
such that the relation $s=dP/dT>0$ is satisfied\,. At high temperatures $\sigma$ and $P$ increase with
$T$ less rapidly than for the case of ideal gas of massless bosons\hspm\footnote
{
Note, that our model does not take into account possible excitations of bosonic resonances
and neglects the deconfinement effects which become more and more important with increasing temperature.
}.

In the case of strong attraction, $\kappa>1$, the condition $M^2(\sigma)<0$ holds in the interval
$\sigma_1<\sigma<\sigma_2$, where $\sigma_{1,2}$ are the two roots of the equation $M(\sigma)=0$:
\bel{spme}
\sigma_{1,2}=\frac{\ds m}{\sqrt{\ds b}}\left(\kappa\mp\sqrt{\kappa^2-1}\right).
\ee
The real solutions of Eq. (\ref{stabc}) exist only outside of this interval.
One can easily see that the true equilibrium state is $\sigma=\sigma_2$\hspm, and
the root $\sigma=\sigma_1$ corresponds to a maximum of the thermodynamic potential $\Omega=-PV$.
It is interesting to note that the lower branch of $\sigma(T)$ bends before reaching the
line $\sigma_{\rm lim}(T)$ (see the curves for \mbox{$\kappa=1.1$} in Fig.~1). After the system reaches the value $\sigma=\sigma_1$ at temperature $T=T_1$, it will
''roll down'' into the state with a larger pressure at $\sigma=\sigma_2$. At fixed temperature this is only
possible by creating a condensate of bosons with zero momentum. In principle, the system may reach metastable states marked in Fig.~1 by the dashed line up to the cross, but at higher temperatures the allowed states lie on the line $\sigma=\sigma_2$.

The thermodynamic characteristics of the 'mixed'  {states}, where the condensate coexists with the normal
phase, can be found from~\mbox{Eqs.~(\ref{scbd}),~(\ref{bter})}, after substitu\-ting \mbox{$\sigma_0=\sigma_2$}\hspm .
In~par\-ticular, the pressure and the energy density in this phase are given by
\bel{pbec1}
P_{\hsp\rm mix}^{(2)}=\frac{g\hspm\pi^2}{90}T^4+P_{\rm ex}(\sigma_2),~~\varepsilon_{\hsp\rm mix}^{(2)}=
\frac{g\hspm\pi^2}{30}T^4-P_{\rm ex}(\sigma_2).
\ee
As can be shown by using~Eqs.~(\ref{skin2}),~(\ref{pbec1}), at \mbox{$T\to 0$} the pressure
$P_{\rm mix}^{(2)}\simeq P_{\rm ex}(\sigma_2)$ becomes negative
at \mbox{$\kappa<2/\sqrt{3}\simeq 1.155$}\hspm\footnote
{
At $\kappa>2/\sqrt{3}$, even the bosonic vacuum at $T=0$ becomes unstable with respect to
 {formation of a classical boson field}. Indeed, at such~$\kappa$ the energy density of the condensate,
\mbox{$\varepsilon_{\rm  {SBC}}(T=0)=-P_{\rm ex}(\sigma_2)$}, becomes negative.
This possibility  {is analogous to a spontaneous creation of the vacuum condensate in the linear sigma model.
We do not consider such a possibility in the present paper.}
}.
On the other hand, as discussed above, the normal states without condensate have a positive
pressure at all~$T$. Hence, the mixed phase becomes unfavorable
at low temperatures.

The true transition point between the normal ('liquid-gas') and the mixed phases
occurs at \mbox{critical} temperature $T_c$, which is
found from the Gibbs condition \mbox{$P_{\rm \hsp lg}\hspm (T_c)=P_{\rm mix}^{(2)}(T_c)$}.
At \mbox{$T=T_c$}, the scalar density~$\sigma$ jumps from some thermally-generated value
$\sigma_c$ to a~larger value $\sigma_2$ which contains the {condensate} 
$\sigma_{\rm cond}=\sigma_2-\sigma_c$.
At the same time, at \mbox{$T=T_c$} the boson effective mass drops from \mbox{$M=M\hspm (\sigma_c)$} to zero, see Fig.~2.
As temperature grows above $T_c$, the condensate gets smaller and finally vanishes at $T=T_2$. We call this unusual behaviour as "triangular phase diagram".

\begin{figure}[htb!]
\includegraphics[trim=0 9mm 0 9mm,width=0.48\textwidth]{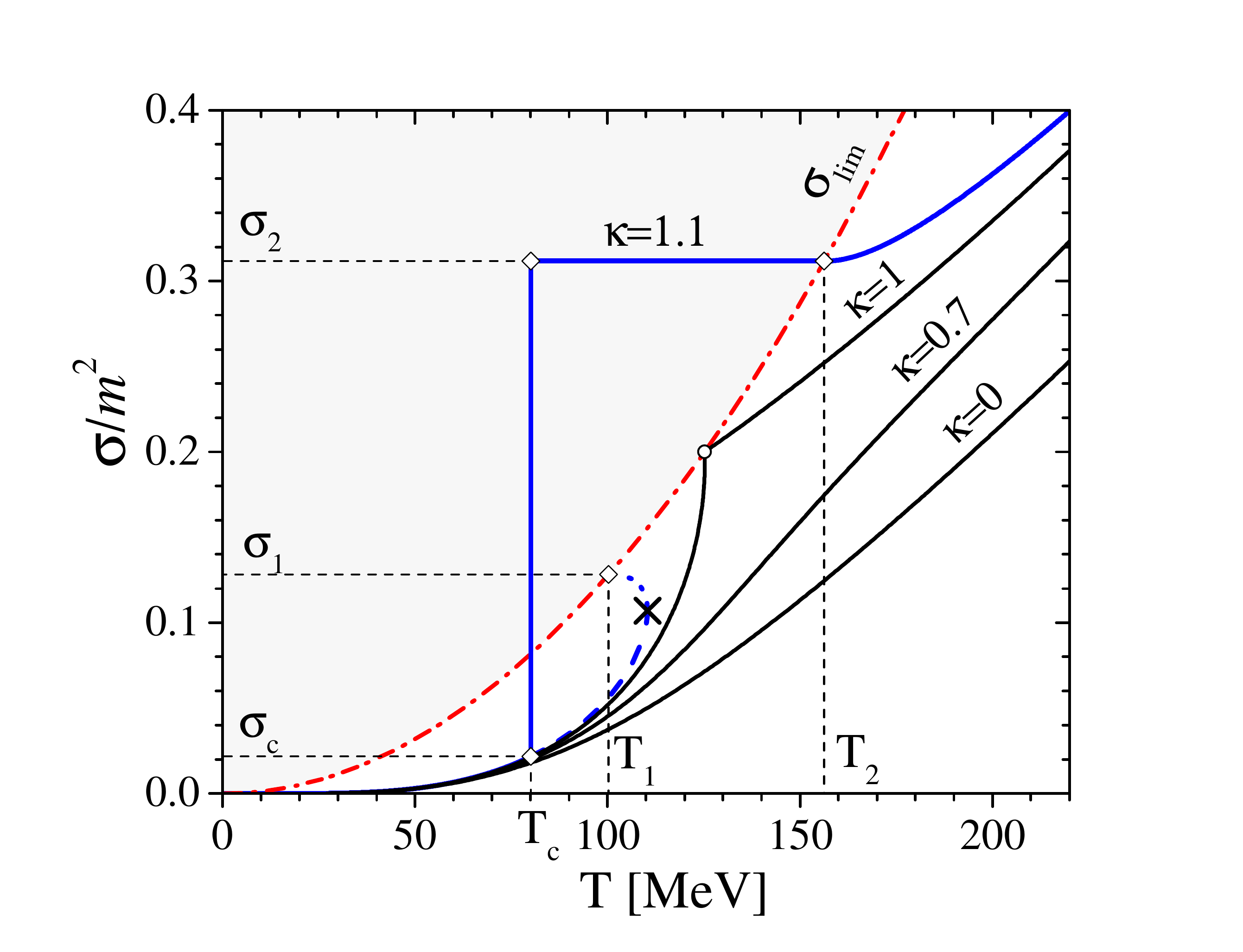}
\caption{
The temperature dependence of the scalar density (in units of~$m^2$)
is shown for several values of the parameter $\kappa$ (the solid lines).
The dash-dotted line shows $\sigma_{\rm lim}$, as defined in~\re{scdm}.
The  {SBC} exists on the horizontal line $\sigma=\sigma_2$,
in the temperature interval from $T_c$ to $T_2$.
The long-dashed line shows the metastable states of the normal phase, while the dotted
line corresponds to unstable states. The cross marks the boundary between
these states.
}
\label{fig-1}
\end{figure}

\begin{figure}[htb!]
\includegraphics[trim=0 9mm 0 9mm,width=0.48\textwidth]{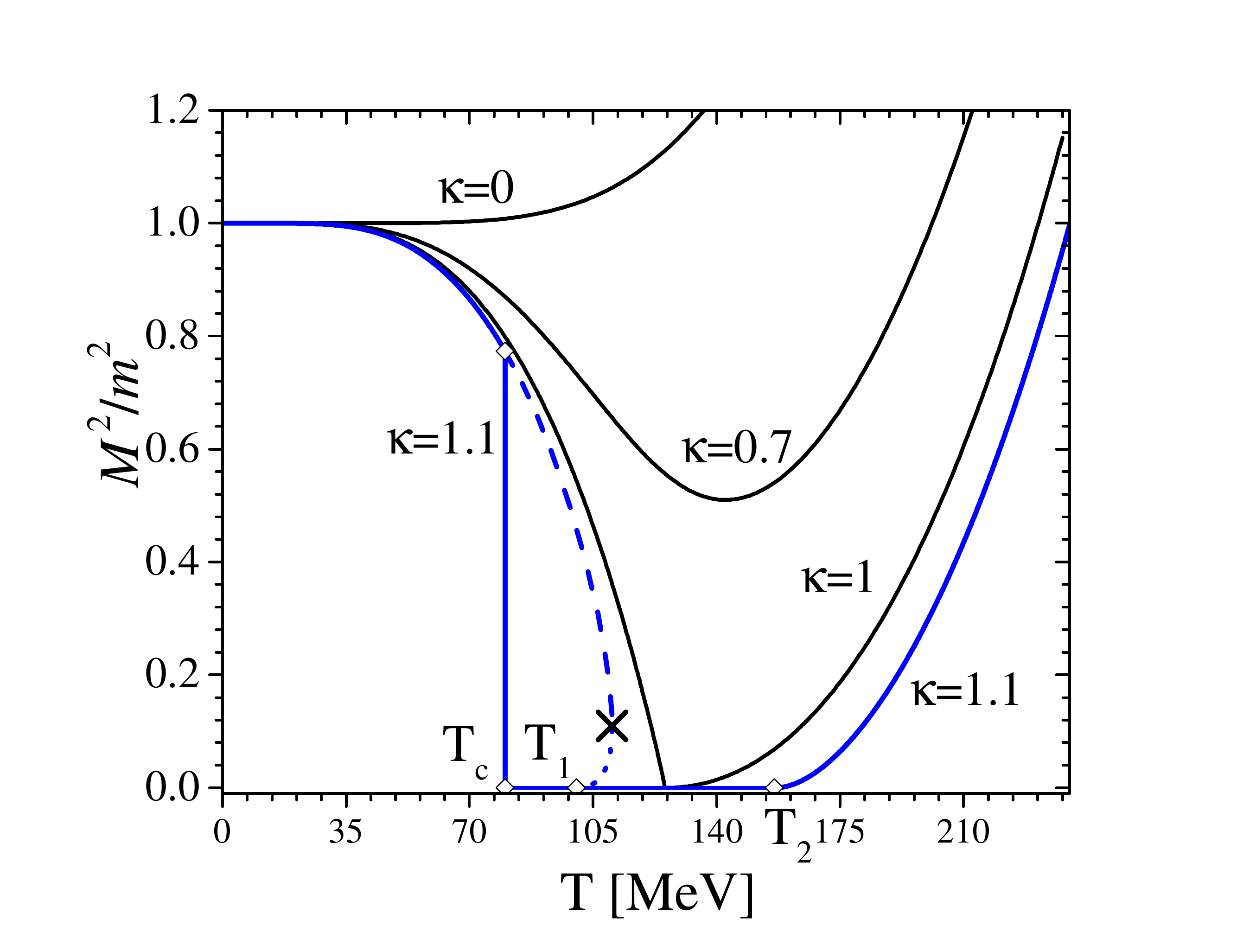}
\caption{
Temperature dependence of the boson effective mass squared (in units of $m^2$) for
several values of $\kappa$. In the supercritical case $\kappa=1.1$ the  {SBC} occurs
in the interval \mbox{$T_c<T<T_2$} where $M^2=0$.
The cross marks the boun\-dary between metastable and unstable states.
}\label{fig-2}
\end{figure}

\section{Numerical results}

\begin{figure}[htb!]
\includegraphics[trim=0 9mm 0 9mm,width=0.48\textwidth]{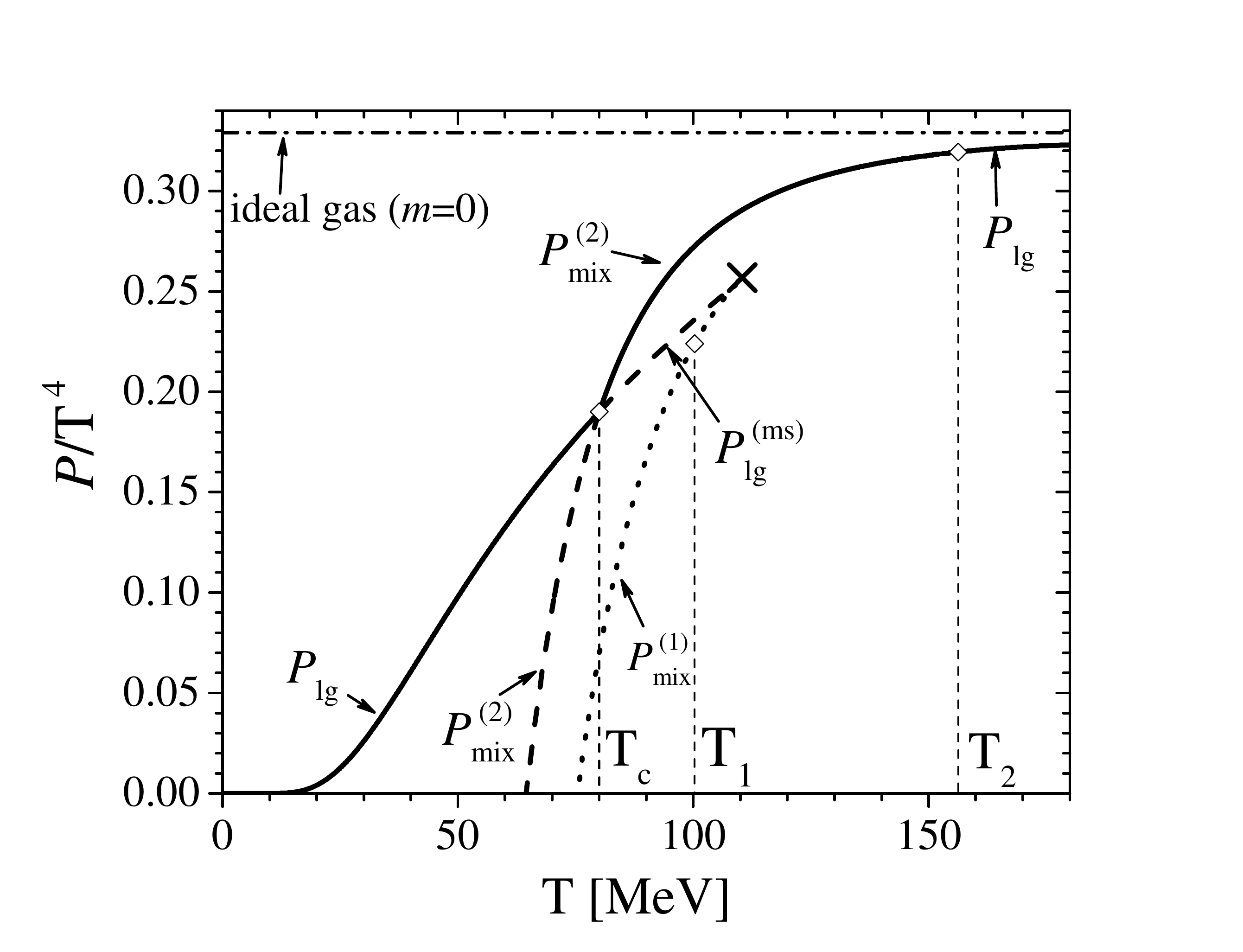}
\caption{
Pressure (in units of $T^4$) as a function of tempe\-rature for the supercritical
case $\kappa=1.1$. The solid, dashed and dotted lines correspond to stable,
metastable and unstable states, respectively. The pressure lines $P^{(1)}_{\rm mix}$
and~$P^{(2)}_{\rm mix}$ are obtained for fixed scalar densities $\sigma=\sigma_1$ and
$\sigma=\sigma_2$, respectively.
The dashed-dotted line corresponds to the ideal gas of massless bosons.
}\label{fig-3}
\end{figure}
The general formalism presented above can be used for any (pseudo)scalar particles with strong self-interaction.~As~an~illustrative example,
we consider pion-like particles with~\mbox{$m=m_{\pi}=140~\textrm{MeV}$}, \mbox{$g=3$}, \mbox{$b=25\hsp m^{-2}$~\cite{anch-2019}}.~Figures~1--3 show our nume\-rical results
for several \mbox{$\kappa$ values}, namely, \mbox{$\kappa=0,0.7,1.0$}~and~$1.1$.~In~the~\mbox{latter} case,
\mbox{$\sigma_2\simeq 0.312\,m^2$} and the  {SBC} phase \mbox{appears} in the temperature
interval \mbox{$T_c<T<T_2$}, where \mbox{$T_2=\sqrt{12\hspm\sigma_2/g}\simeq 156~\textrm{MeV}$} and
\mbox{$T_c\simeq 80~\textrm{MeV}$} is the critical temperature of the first-order PT.
At this tempe\-rature, the scalar density jumps from \mbox{$\sigma_c\simeq 0.022\,m^2$} to~$\sigma_2$
and then remains constant until $T=T_2$. At higher temperatures the condensate disappears, and
the scalar density $\sigma$ is generated entirely by thermal excitations (see the branch starting
from $(T_2,\sigma_2)$ in Fig.~1).

As demonstrated above, the phase transition predicted by the model (for $\kappa>1$) is rather unusual.
With increasing temperature it starts at some temperature $T=T_c$ as the first-order transition with
a jump of scalar density from $\sigma_c$ to $\sigma_2$. With further increase of temperature
the scalar density remains constant, $\sigma=\sigma_2$, but the condensate density
$\sigma_{\rm cond}=\sigma_2-\sigma_{\rm th}$ decreases and finally vanishes at $T=T_2$ where
$\sigma_{\rm th}=\sigma_2$ . Therefore, we do not expect that higher-order
fluctuations will destroy or significantly modify the behaviour of the scalar density at $T\simeq T_c$\hsp .
On the other hand, at $T=T_2$ the condensate vanishes smoothly and only thermal fluctuations
remain at $T>T_2$\hsp .

Finally we would like to point out that pions have a~special nature as Goldstone bosons of
spontaneously broken chiral symmetry, and by this reason require a~special treatment, like that in
the linear sigma-model~\cite{Len00}.  In this case
the scalar condensate exists already in the vacuum at T=0 but melts at higher temperatures.
In the future we are planning to study the mesonic sector of this model in more details.

\section{Concluding remarks}

In this paper we have presented a thermodynamically
consistent model to describe dense bosonic systems at high temperatures and zero chemical potential.
A central step of our approach is to solve
Eq.~(\ref{gape1}) for the boson scalar density~$\sigma$ as a function of
temperature. We show that if the attractive mean field is so strong that the stability
condition $M^2(\sigma)>0$ is violated,
the classical scalar field (condensate) forms in the multi-boson system.
Our analysis leads to the conclusion that in the presence of a condensate, the
allowed states of the system satisfy the condition $M\hspm (\sigma)=0$, i.e.,
the bosonic quasiparticles  {are} massless.

It is known that loop corrections and higher-order fluctuations may change
the second-order phase transition to first order or crossover, see e.g.~\cite{Kap06,Kad09}.
As far as we know, there are no dedicated studies for the $\phi^4-\phi^6$ model considered here.
However, we do not think that the above-mentioned corrections can qualitatively change the character
of a strong first-order phase transition predicted by our calculations.

Of course, at high temperatures considered in this paper
other hadronic degrees of freedom will be present in the system. They  {may} add
additional terms to the effective potential which are proportional to the density of
these species~\cite{Shu91}. These terms will reduce the meson mass~$m$ and
thus, the threshold for Bose condensation (see~\re{efma}).

\section*{Acknowledgements}
The authors thank M.~I.~Gorenstein for useful discussions.
The work of \mbox{D.~V.~A.} is supported by the Programs
"The structure and dynamics of statistical and quantum-field systems"
and "The dyna\-mics of formation of spatially-heterogeneous
structures in many-particle systems" of the Department of Physics and Astro\-nomy of NAS of Ukraine.
\mbox{I.~N.~M.} acknowledges the financial support from the Helmholtz International Center for
FAIR, Germany. \mbox{L.~M.~S.} appreciates the support from the
Frankfurt Institute for Advanced Studies. H.~St. thanks for support from the
J.~M. Eisenberg Professor Laureatus of the Fachbereich Physik.


\end{document}